
\font\rmu=cmr10 scaled\magstephalf
\font\bfu=cmbx10 scaled\magstephalf

\font\it=cmti10 scaled \magstephalf

\rmu

\font\rmus=cmr8
\font\rmuss=cmr6
\font\mait=cmmi10 scaled\magstephalf
\font\maits=cmmi7 scaled\magstephalf
\font\maitss=cmmi7
\font\msyb=cmsy10 scaled\magstephalf
\font\msybs=cmsy8 scaled\magstephalf
\font\msybss=cmsy7
\font\bfus=cmbx7 scaled\magstephalf
\font\bfuss=cmbx7
\font\cmeq=cmex10 scaled\magstephalf

\textfont0=\rmu
\scriptfont0=\rmus
\scriptscriptfont0=\rmuss

\textfont1=\mait
\scriptfont1=\maits
\scriptscriptfont1=\maitss

\textfont2=\msyb
\scriptfont2=\msybs
\scriptscriptfont2=\msybss

\textfont3=\cmeq
\scriptfont3=\cmeq
\scriptscriptfont3=\cmeq

\newfam\bmufam  \textfont\bmufam=\bfu
      \scriptfont\bmufam=\bfus \scriptscriptfont\bmufam=\bfuss

\hsize=15.5cm
\vsize=21cm
\baselineskip=16pt   
\parskip=12pt plus  2pt minus 2pt

\def\a{\alpha}
\def\b{\beta}
\def\d{\delta}
\def\e{\epsilon}

\def\semi{\bigcirc\kern-1em{s}\;}

\def\del{\partial}
\def\ni{\noindent}
\def\R{{\rm I\!R}}

\def\one{{\mathchoice {\rm 1\mskip-4mu l} {\rm 1\mskip-4mu l}
{\rm 1\mskip-4.5mu l} {\rm 1\mskip-5mu l}}}
\def\Q{{\mathchoice
{\setbox0=\hbox{$\displaystyle\rm Q$}\hbox{\raise 0.15\ht0\hbox to0pt
{\kern0.4\wd0\vrule height0.8\ht0\hss}\box0}}
{\setbox0=\hbox{$\textstyle\rm Q$}\hbox{\raise 0.15\ht0\hbox to0pt
{\kern0.4\wd0\vrule height0.8\ht0\hss}\box0}}
{\setbox0=\hbox{$\scriptstyle\rm Q$}\hbox{\raise 0.15\ht0\hbox to0pt
{\kern0.4\wd0\vrule height0.7\ht0\hss}\box0}}
{\setbox0=\hbox{$\scriptscriptstyle\rm Q$}\hbox{\raise 0.15\ht0\hbox to0pt
{\kern0.4\wd0\vrule height0.7\ht0\hss}\box0}}}}
\def\C{{\mathchoice
{\setbox0=\hbox{$\displaystyle\rm C$}\hbox{\hbox to0pt
{\kern0.4\wd0\vrule height0.9\ht0\hss}\box0}}
{\setbox0=\hbox{$\textstyle\rm C$}\hbox{\hbox to0pt
{\kern0.4\wd0\vrule height0.9\ht0\hss}\box0}}
{\setbox0=\hbox{$\scriptstyle\rm C$}\hbox{\hbox to0pt
{\kern0.4\wd0\vrule height0.9\ht0\hss}\box0}}
{\setbox0=\hbox{$\scriptscriptstyle\rm C$}\hbox{\hbox to0pt
{\kern0.4\wd0\vrule height0.9\ht0\hss}\box0}}}}

\font\fivesans=cmss10 at 4.61pt
\font\sevensans=cmss10 at 6.81pt
\font\tensans=cmss10
\newfam\sansfam
\textfont\sansfam=\tensans\scriptfont\sansfam=\sevensans\scriptscriptfont
\sansfam=\fivesans
\def\sans{\fam\sansfam\tensans}
\def\Z{{\mathchoice
{\hbox{$\sans\textstyle Z\kern-0.4em Z$}}
{\hbox{$\sans\textstyle Z\kern-0.4em Z$}}
{\hbox{$\sans\scriptstyle Z\kern-0.3em Z$}}
{\hbox{$\sans\scriptscriptstyle Z\kern-0.2em Z$}}}}

\newcount\foot
\foot=1
\def\note#1{\footnote{${}^{\number\foot}$}{\ftn #1}\advance\foot by 1}

\def\frac#1#2{{#1\over #2}}
\def\text#1{\quad{\hbox{#1}}\quad}

\font\ch=cmbx12 scaled\magstephalf
\font\ftn=cmr8 scaled\magstephalf

\font\it=cmti10 scaled\magstephalf

\font\titch=cmbx12 scaled\magstep2
\font\titname=cmr10 scaled\magstep2
\font\titit=cmti10 scaled\magstep1
\font\titbf=cmbx10 scaled\magstep2

\nopagenumbers

\line{\hfil DFF 244/02/96}
\line{\hfil February 21, 1996}
\vskip2.2cm
\centerline{\titch A REAL ALTERNATIVE TO QUANTUM }
\vskip.5cm
\centerline{\titch GRAVITY IN LOOP SPACE}
\vskip1.7cm
\centerline{\titname R. Loll\note{Supported by the European Human
Capital and Mobility program on ``Constrained Dynamical Systems"}}
\vskip.5cm
\centerline{\titit Sezione INFN di Firenze}
\vskip.2cm
\centerline{\titit Largo E. Fermi 2}
\vskip.2cm
\centerline{\titit I-50125 Firenze, Italy}

\vskip3.5cm
\centerline{\titbf Abstract}
\vskip0.7cm
We show that the Hamiltonian of four-dimensional Lorentzian
gravity, defined on a space of real, SU(2)-valued connections,
in spite of its non-polynomiality possesses a natural quantum 
analogue in a lattice-discretized formulation of the theory.
This opens the way for a systematic investigation of its 
spectrum. The unambiguous and well-defined scalar product is
that of the SU(2)-gauge theory.
We also comment on various aspects of the continuum theory.

\vfill\eject
\footline={\hss\tenrm\folio\hss}
\pageno=1





In spite of considerable advances in our understanding of the
canonical quantization of gravity, based on the use of complex
connection variables [1] and quantum representations on spaces
of Wilson loops [2], some basic problems have remained
unsolved. Firstly, despite the apparent simplicity and polynomiality of
the Hamiltonian in this approach, even at a formal level only a few
(and for the most part physically uninteresting) 
solutions to the Wheeler-DeWitt equation have been found in any
loop representation. Secondly, treatment of the reality conditions 
(which have to be imposed on the $SL(2,\C)$-valued basic variables to
recover real Einsteinian gravity) continues to be troublesome in
the quantum theory. (The heat kernel measure on
$SL(2,\C)$-wave functions used in [3,4] provides a 
kinematic scalar product for complex Wilson loops (modulo 
a possible hidden metric dependence), but does not incorporate 
correctly the reality conditions for Ashtekar gravity.)

Faced with these difficulties, it may be time to remember that there
exists a version of Hamiltonian gravity in terms of {\it real}
connection variables 
[5] (for a corresponding action principle, see
[6]),  and to re-evaluate the achievements and drawbacks of the
complex formulation as compared to this real alternative. There
is a ``unified" derivation of the two connection representations 
in the classical theory: 
starting from the 3+1 formulation in terms of a cano\-ni\-cally
conjugate pair $(P_i^a,K_a^i)$ of $SO(3)$-valued variables, where
$P_i^a$ denotes a dreibein (with density weight $1$) and 
$K_a^i=\frac{1}{\sqrt{g}}K_{ab} P^{bi}$ is (whenever the Gauss law
constraints are satisfied) the
extrinsic curvature with one index raised, one may define a canonical
transformation 

$$
\eqalign{
&E_i^a=\alpha P_i^a\cr
&A_a^i=\Gamma_a^i +\beta K_a^i,}\eqno(1)
$$

\ni where $\Gamma\equiv\Gamma(P)$ is the spin connection compatible
with $P$, and $\alpha$ and $\beta$ are two non-vanishing constants.
In terms of the new variables, one has

$$
\{A_a^i(x),E_j^b(y)\}=\alpha\beta\{ K_a^i(x),P_j^b(y)\}=-\alpha\beta
\delta^i_j \delta_a^b \delta^3 (x-y).
$$

\ni The Hamiltonian constraint function in terms of $(A,E)$ reads

$$
\epsilon^{ijk} E_i^a E_j^b F_{ab}^k -(\frac{2}{\beta^2}+2)
E_i^{[a} E_j^{b]} (A_a^i-\Gamma_a^i)(A_b^j-\Gamma_b^j).
\eqno(2)
$$

\ni Ashtekar's choice is $\a=1$, $\b=-i$, which makes the second,
non-polynomial term drop out of the Hamiltonian. The drawback
is that according to formula (1), the connection
variable $A_a^i$ is now complex. Our real version of connection
gravity consists in choosing $\alpha=1$ and $\beta=-1$ [5].
The functional forms for the remaining spatial diffeomorphism
constraints $E_k^a F_{ab}^k=0$ and Gauss law constraints
$\nabla_a E^a_i=0$ are independent of the choice of 
$\alpha$ and $\beta$. (As usual, we denote the field strength of
the connection $A_a^i$ by $F$ and its covariant derivative by
$\nabla$.) Their treatment therefore does not have
to be changed with respect to the usual, complex formulation.
In particular, we will continue to use Wilson loop variables
in the quantum theory. 
In the real formulation, (2) becomes

$$
H^{\R}=\epsilon^{ijk} E_i^a E_j^b F_{ab}^k -H^{\rm pot},
\eqno(3)
$$

\ni retaining a ``potential" term 
(which is a misnomer since it depends both on coordinates and
momenta). Starting from a form equivalent to (2) (formula (14) in
Barbero's paper [5]) for $H^{\rm pot}$, one finds after
some algebra that it may be re-expressed 
as a polynomial in $A$ and $E$, up to determinantal factors, namely,

$$
\eqalign{
H^{\rm pot}=& (\det E)^{-2} \eta_{a_1 a_3 a_4}\eta_{b_1 b_3 b_4}
(E^{a_3}_k E^{a_4}_l E^{b_3}_m E^{b_4}_n\cr
&- 2 E^{a_3}_m E^{a_4}_n E^{b_3}_k E^{b_4}_l) 
E^{a_2}_k E^{b_2}_m 
(\nabla_{a_2} E^{a_1}_l)(\nabla_{b_2} E^{b_1}_n),}\eqno(4)
$$

\ni and up to terms proportional to the Gauss law constraints.

Let us now recall some features of the complex formulation
with the Hamiltonian $H^{\C}=\epsilon^{ijk} E_i^a E_j^b F_{ab}^k$.
There exist simple solutions to $\hat H^{\C}\Psi=0$ in the
loop representation, both in the formal continuum approach [2] 
and on the lattice [4], where the loop state $\Psi$ depends on 
(smooth) non-intersecting loops 
(see also [7] for a generalization within a lattice language). 
This is a straightforward consequence of the
antisymmetry of $H^\C$ in the spatial indices $a$ and $b$. However,
these solutions are probably not interesting from a physical
point of view, because they
correspond to zero-eigenstates of the volume operator
[8,9]. We are not aware of any non-trivial (in this
sense) solutions that have been found by tackling the equation $\hat
H^{\C}\Psi=0$ directly.  

Other interesting features of the complex formulation are that all
four diffeomorphism constraints can be solved by making a
so-called Capovilla-Dell-Jacobson ansatz [10], and that there
exists a formal solution to the Wheeler-DeWitt equation (with
a specific factor ordering for $\hat H^{\C}$), given by the
exponential of the Chern-Simons action for the complex connection
$A$ [11]. However, in the absence of a proper treatment of the
reality conditions, the significance of these properties for the full
gravitational theory has remained unclear. There is a version of the
CDJ-ansatz in the real theory [12], but the remaining 
Gauss law constraints are considerably more difficult. 
The real Chern-Simons term does not seem to play a special role
(apart from being a generating functional for the $B$-field,
$B^{ai}=\frac{1}{2}\eta^{abc} F_{bc}^i$, like in the complex theory).
This also implies that the solutions to the Hamiltonian related to
knot invariants (obtained by a formal loop transform of the 
Chern-Simons term from the connection to the loop representation 
[13]) do not carry over to the real theory. The absence of
these ``nice'' features from the real theory may lead one to 
wonder whether they may not go away also in the complex case once
reality conditions are properly taken into account. 

The (quantum) Hamiltonian $H^{\C}$ of the complex theory has a natural
representation in terms of loop functions, since the components
of the field strength $F$ both in the continuum [2] and on
the lattice [16] can be obtained by considering infinitesimal
planar Wilson loops.  We will show below that a similar statement
holds for the real Hamiltonian $H^{\R}$, i.e. in spite of its
non-poly\-nomiality, the potential term has a natural representation on
quantum loop states, at least in the lattice formulation. This makes
the search for zero eigenvectors of $\hat H^{\R}$ accessible
numerically. Clearly the potential term presents a computational
complication, which however in the absence of any explicit solutions of
$\hat H^{\C}\Psi=0$ seems to be a matter of degree rather than one of
principle. The big advantage of the real formulation is the
presence of a well-defined and unambiguous scalar product on Wilson 
loop functionals in the quantum theory, that induced by the Haar
measure of SU(2).

The following discussion will take place within the discrete
lattice framework, with occasional comments on a possible
continuum formulation. Furthermore, we will focus on the discussion
of the potential term, which is new with respect to previous
treatments [16,4,7]. It may be worthwhile noticing 
that in the lattice approach, both the state space and the operators
get regulated simultaneously, since they share the same support (in 
terms of lattice links); in discretizing the state space, we get
a regularization of the Hamiltonian ``for free''.

Recall now the basic ingredients of the Hamiltonian lattice
formulation for theories based on a space of connections
[17]. Our lattice will be a cubic $N\times N\times
N$-lattice, with periodic boundary conditions. The basic operators
associated with each lattice link $l$ are in our case an $SU(2)$-link
holonomy $\hat V$ (represented by multiplication by $V$), 
together with its inverse $\hat V^{-1}$, and a pair
of canonical momentum operators $\hat p^+_i$ and $\hat p^-_i$, where
$i$ is an adjoint index. The operator $\hat p^+_i(n,\hat a)$ is based
at the vertex $n$, and is associated with the link $l$ oriented in the
positive $\hat a$-direction. By contrast, $\hat p^-_i(n+\hat 1_{\hat
a},\hat a)$ is based at the vertex displaced by one lattice unit in
the $\hat a$-direction, and associated with the inverse link
$l^{-1}(\hat a)=l(-\hat a)$. The wave functions are elements of
$\times_l L^2(SU(2),dg)$, with the product taken over all links, and
the Haar measure $dg$. The basic commutators are 

$$
\eqalign{
&[\hat V_A{}^B(n,\hat a),\hat V_C{}^D(m,\hat b)]=0\cr
&[\hat  p^+_i(n,\hat a),\hat V_A{}^C(m,\hat b)]=
-\frac{i}{2}\,\d_{nm}\d_{\hat a\hat b}\, \tau_{iA}{}^B\hat V_B{}^C
\cr
&[\hat  p^-_i(n,\hat a),\hat V_A{}^C(m,\hat b)]=
-\frac{i}{2}\,\d_{nm}\d_{\hat a\hat b}\,\hat V_A{}^B \tau_{iB}{}^C
\cr
&[\hat p^\pm_i(n,\hat a),\hat p^\pm_j(m,\hat b)]=
\pm i\, \d_{nm}\d_{\hat a\hat b}\, \e_{ijk}\, \hat p_k\cr
&[\hat p^+_i(n,\hat a),\hat p^-_j(m,\hat b)]=0,}\eqno(5)
$$

\ni where $\e_{ijk}$ are the structure constants of $SU(2)$. Our
normalization for the $SU(2)$-generators $\tau_i$ is such that
$[\tau_i,\tau_j]=\epsilon_{ijk}\tau_k$ and $A_a=A_a^i \tau_i/2$.
Taking into account the expansions 

$$
\eqalign{
&V_A{}^B(\hat b)=\one_A{}^B+a\, A_{bA}{}^B+O(a^2)\cr
&p_i^\pm (\hat b)=a^2 E^b_i +O(a^3)}\eqno(6)
$$

\ni of the corresponding classical quantities for 
small lattice spacing $a$, one derives the following expansion

$$
\eqalign{
-\frac{1}{2}{\rm Tr}\,(\tau_i &V(n,\hat b)p^\pm_j(n+\hat 1_{\hat
b},\hat c) \tau_j V(n,\hat b)^{-1})-p^\pm_i(n,\hat c)=\cr
&  a^3 (\del_b E_i^c+ \epsilon_{ijk}A_{bj} E_k^c)+O(a^4)=a^3
\nabla_b E_i^c+O(a^4).}\eqno(7)
$$
  
\ni The prefactor $-1/2$ occurs because of Tr$\,\tau_i\tau_j =
-2\delta_{ij}$. 

For reasons of symmetry we will from now
on use the averaged momenta $p_i:=(p_i^+ +p_i^-)/2$ and their quantum
versions. Motivated by (7), we may represent the quantum
covariant derivative $\hat{\nabla_b E_i^c}$ by the lattice operator 

$$
-\frac{1}{2}{\rm Tr}\,(\tau_i V(n,\hat b)\hat p_j(n+\hat 1_{\hat
b},\hat c) \tau_j V(n,\hat b)^{-1})-\hat p_i(n,\hat c).\eqno(8)
$$

\ni Note, however, that this operator is well defined only on those
Wilson loop states that for each occupied link
$l(n,\hat c)$ have also the neighbouring ``parallel" link
$l(n+\hat 1_{\hat b},\hat c)$ occupied. (A lattice Wilson loop 
is a gauge-invariant function of the form Tr$\,V(l_1)V(l_2)...
V(l_k)$, with $\gamma=l_1\circ l_2\circ ...\circ l_k$ a closed
loop of lattice links.)

This happens because the left-hand side of expression 
(7) is a finite difference. If one of
$\hat p_j(n+\hat 1_{\hat b},\hat c)$ and $\hat p_i(n,\hat c)$
but not the other vanishes on a state $\Psi$, 
the result of the action of (8) on $\Psi$ 
for small lattice spacing
$a$ is of lower order in $a$ and diverges in the limit as
$a\rightarrow 0$. In particular, loop states with only sparse
intersections have a singular behaviour under the action of the
covariant derivative operator (8). 
Moreover, since we do not want to distinguish any
particular direction on the lattice, we will instead of (8)
use the averaged version

$$
\eqalign{
-\frac{1}{4}({\rm Tr}\,(\tau_i V(n,\hat b)\hat p_j&(n+\hat 1_{\hat
b},\hat c) \tau_j V(n,\hat b)^{-1})\cr
&-{\rm Tr}\,(\tau_i V(n-\hat 1_{\hat b},\hat b)^{-1} \hat
p_j(n-\hat 1_{\hat b},\hat c) \tau_j V(n-\hat 1_{\hat b},\hat b))).}
\eqno(9)
$$
\ni A remark similar to the one made 
above concerning the allowed loop states
applies to this operator as well. 

To obtain a well-defined lattice operator $\hat H^{\rm pot}$,
we still must take care about the determinantal factor
$(\det E)^{-2}$. As shown in [9,14], 
$\det E=\frac{1}{3!} \eta_{abc}\,\e^{ijk} E^a_i E^b_j  E^c_k$
possesses a natural quantum lattice analogue 
$\hat D(n):=\frac16 \eta_{abc}\,\e^{ijk} \hat p_i(n,\hat a) 
\hat p_j(n,\hat b) \hat p_k(n,\hat c)$. Since the latter is a
selfadjoint operator, there exists a Hilbert space basis in which
it is diagonal. For the gauge-invariant sector,
this basis is most easily constructed in terms of so-called spin
network states, certain (anti-)symmetrized,
real linear combinations of Wilson loop states.  

A spin network
associates a positive ``occupation number" with each lattice link,
which may be interpreted as
counting the number of (unoriented) flux lines of basic spin-$\frac12$
representations along the link, and also keeps track of the way in
which those flux lines can be contracted gauge-invariantly at the
vertices (see [18] for more details). A concrete way of
constructing elements of the spin network basis is to begin with
sets of Wilson loops with fixed 
occupation numbers and arbitrary intertwiners, and
then select linearly independent sets of intertwiners  at the vertices
(which generally, in terms of a loop language, 
are still related by Mandelstam constraints).

In terms of such states, the diagonalization of the operators
$\hat D(n)$ is reduced to the diagonalization within finite-dimensional
subspaces of the Hilbert space. In the resulting diagonal basis we can 
meaningfully define quantum representations of arbitrary functions
of $(\det E)$ in terms of their eigenvalues. (Investigations of
the spectrum of (two related but not identical versions of) 
the volume operator have been performed in [14,15].) In
particular, if we restrict ourselves to eigenstates with
non-vanishing eigenvalues, we can quantize $(\det E)^{-2}$ on
the lattice. There is no immediate analogue of this construction in
the continuum, although one can define a quantized version
of the classical volume function $\int \sqrt{| \det E |}$
[8]. Since the naive local quantum operator $\hat{\det E}$
vanishes at all points of a loop state without intersections, 
$1/\hat{\det E}$ is ill-defined almost everywhere on a typical
loop state. In the continuum there thus seem to be no good analogues
of the lattice states with ``volume everywhere".  A way
out of this may be to either use smeared-out wave functions and/or
regularize the quantum operators appropriately. Alternatively, one
may multiply the Hamiltonian $H^\R$ by a factor $(\det E)^2$.
Classically, this changes the constraint algebra at most by
terms proportional to the constraints. In particular, the commutator
of two Hamiltonians is just rescaled by a factor of
$(\det E)^4$. Quantum-mechanically, however, operator ordering problems
may appear. 

Possibly the problem is not as serious as it seems since along smooth
pieces of loop where $\hat{\det E}$ vanishes, also $\hat{ (\det
E)^2}\hat H^{\rm pot}$ does, and one may be able to regularize the
action of $\hat H^{\rm pot}$ to a finite value. Another problem, 
also relevant to the lattice approach, is that $\hat{\det E}$
has many zero-volume eigenstates even at loop intersections
[9,14]. Hence the question is whether one can
consistently restrict the Hilbert space so that the action of $\hat
H^{\rm pot}$ is always well defined (recall that $[\hat H^\R,\hat{\det
E}]\not=0$). For example, no immediate problems arise if one chooses
a factor ordering for $\hat H^{\rm pot}$ with the $\det E$-terms
to the right and discards all zero-volume states by hand. 

To summarize: there is a well-defined regularized, self-adjoint operator 
expression for the real Hamiltonian (3) on the lattice, 
at least on a large subsector of
the Hilbert space of gauge-invariant functions. This operator is
subject to the usual ambiguities with regard to factor ordering and
addition of higher-order terms in the lattice spacing
$a$. For example, we may choose

$$
\eqalign{
&\hat H^{\R,latt}(n)=-\epsilon_{ijk}{\rm Tr}\,(\hat V(n,P_{\hat a
\hat b})\tau_k)\;\hat p_i(n,\hat a)\hat p_j(n,\hat b)\cr
&\;-\eta_{a_1 a_3 a_4}\eta_{b_1 b_3 b_4}(\,\hat p_k (n,\hat a_3)
\hat p_l(n,\hat a_4) \hat p_m(n,\hat b_3)\hat p_n(n,\hat
b_4)\cr
&\hskip2cm - 2 \hat p_m (n,\hat a_3) \hat p_n(n,\hat a_4)
\hat p_k(n,\hat b_3)\hat p_l(n,\hat b_4)\, )\;\hat p_k(n,\hat a_2)\hat
p_m(n,\hat b_2) \times\cr
&\;\,\frac{1}{4}{\rm Tr}(\,
\tau_l\hat V(n-\hat 1_{\hat a_2},\hat a_2)^{-1}
\hat p_s(n-\hat 1_{\hat a_2},\hat a_1)\tau_s \hat 
V(n-\hat 1_{\hat a_2},\hat a_2)\cr
&\hskip3cm -\tau_l\hat V(n,\hat a_2)\hat p_s(n+\hat
1_{\hat a_2},\hat a_1)\tau_s \hat V(n,\hat a_2)^{-1}
)\times\cr
&\hskip0.5cm {\rm Tr}(\,
\tau_n\hat V(n-\hat 1_{\hat b_2},\hat b_2)^{-1}
\hat p_t(n-\hat 1_{\hat b_2},\hat b_1)\tau_t \hat 
V(n-\hat 1_{\hat b_2},\hat b_2)\cr
&\hskip3cm -\tau_n\hat V(n,\hat b_2)\hat p_t(n+\hat
1_{\hat b_2},\hat b_1)\tau_t \hat V(n,\hat b_2)^{-1}
 )\;\hat D(n)^{-2}}\eqno(10)
$$

\ni for the Hamiltonian localized around a vertex $n$. In (10),
all spatial indices $\hat a$, $\hat b$ etc. are summed over,
and $V(n,P_{\hat a\hat b})$ is the holonomy associated with
a plaquette loop in the $\hat a$-$\hat b$-plane. 

Note that the Hamiltonian of metric gravity (also containing inverse
powers of $\det g=|\det E|$) cannot be treated in a similar way.
The construction above depended on (i) the reformulation
of canonical gravity in terms of connection variables, hence (ii) the 
possibility of choosing a gauge-invariant Hilbert space of Wilson
loops, therefore (iii) the diagonalization of the operators
$\hat D(n)$ in terms of spin network states, together with (iv) a
natural regularization of the covariant derivative terms in $\hat 
H^{\rm pot}$. 

Having thus set the stage for a systematic investigation of
the Hamiltonian eigenvalue problem $\hat H^{\R,latt}\Psi=0$, we will
now describe some technical problems that have to be addressed for
its solution. 
Consider the action of a local lattice Hamiltonian $\hat
H^{\R,latt}(n)$ on a spin network state $\Psi$. Since the
momentum operators do not change the occupation
numbers $j_i$ of links, it {\it a priori} looks as if this action would
result in a set of loop states with $\Delta j_i\in\{ 0,+1,+2\}$,
depending on the contributions $\hat V(link)$ to the various links in a
neighbourhood of $n$, coming from the kinetic and potential parts of
the Hamiltonian. Unfortunately, life is not as simple. 

Take, for example,
the action of the polynomial part $\hat O$ of $\hat H^{\rm pot}$ on a
spin network state. Our computations show
that the resulting terms generically do not form
a set of states that combine in a simple way to give one (or a small
number of) spin networks, because the operator action does not
preserve the total (anti-)symmetry over link permutations of the
spin network. The result must always be expressible as a unique linear
combination of spin network states, but it turns out that this
decomposition in general contains
states whose occupation numbers differ from those of the original
state $\Psi$ within a whole range of values. For instance, the action of
$\hat O(n)$ on a state $\Psi$ with occupation number $j_i$ for some link
$l_i$ based at the vertex $n$ may result in a sum of spin network
states with $j_i$'s taking any one of the positive values
$j_i+4,j_i+2,j_i,j_i-2,\dots$. Moreover, through retracings of
the form $V(n',\hat a)V(n',\hat a)^{-1}\equiv \one$ occurring
during the decomposition of $\hat O\Psi$ into independent spin
networks, even links may be affected that where not acted on
directly by $\hat O(n)$ in the first place. This reveals a somewhat
unpleasant property of the spin network states which in a sense are
``maximally non-local" (as opposed to sets of maximally
localized Wilson loop functions that one may favour
in certain gauge-theoretic applications [19]), especially in 
conjunction with our requirement of selecting only spin
networks with non-zero volume at every vertex. 

We therefore conclude that the investigation of the spectrum of $\hat
H^{\R,latt}$ requires the pre\-sence of an efficient algorithm for
generating independent spin network states and computing inner
products of such states. 
We reckon that even in the discretized
lattice version the spectral problem is sufficiently complicated so
as to make further approximations necessary. Since we can
calculate matrix elements of the Hamiltonian explicitly, we can
neglect small contributions, depending on suitable perturbation
parameters such as those characterizing the spin network states or
related to the bare gravitational coupling constant $G$. 
More details on our investigation of these issues will appear
elsewhere.

Coming back to the continuum theory, one can show that the
linearized limit of the real connection formulation 
coincides with that of the complex one (c.f. the treatment of
usual Ashtekar gravity in [20]). In the real case,
the linearized versions of the kinetic and potential terms in
(3) become proportional to each other and add up to the
expected result. Likewise, the large-$G$ limit, as, for example,
discussed in [21], is unchanged. For the $G\rightarrow 0$
limit [22], this does not necessarily seem to be the case.

Our real treatment (or an appropriate continuous analogue)
can be viewed as complementary to another
approach that has recently been suggested for dealing with the
complex version of the theory, namely, making use of a ``generalized
Wick transform" [23]. In this ansatz, one tries to define a
transformation $\hat W=\exp\hat C$, with $C=\frac{\pi}{2}\int K_a^i
E_i^a$, between two Hilbert spaces where in one the Hamiltonian has
the simple form  $H^{E}=\epsilon^{ijk} E_i^a E_j^b F_{ab}^k$
and in the other the more complicated form of the real
theory. As far as we understand, the difficulties in making the operator $\hat
W$ well defined in the continuum quantum theory are roughly comparable
to those of constructing the continuum Hamiltonian $\hat H^\R$.
(Note that, like the phase space functional $C$, 
also the potential term $H^{\rm pot}$ can be written in terms of
Poisson commutators of the quantities $\int (\det E)^{-\frac{1}{2}}
H^E$, $\int\sqrt{ \det E }$, $A_a^i$ and $E_i^a$.) However, even if
these could be overcome, the problem of finding non-trivial solutions
to the Wheeler-DeWitt equation would still remain. 

\ni{\it Acknowledgement.} The author is indebted to F. Barbero for
numerous discussions on real connection gravity.

\vskip2cm
\vfill\eject

\line{\ch References\hfil}

\item{[1]} A. Ashtekar: New variables for classical and quantum
gravity, {\it Phys. Rev. Lett.} 57 (1986) 2244-7; A new 
Hamiltonian formulation of general relativity, {\it Phys. Rev.} D36
(1987) 1587-1603

\item{[2]} C. Rovelli and L. Smolin: Loop space representation of
quantum general relativity, {\it Nucl. Phys.} B331 (1990) 80-152

\item{[3]} A. Ashtekar, J. Lewandowski, D. Marolf, J. Mour\~ao
and T. Thiemann: Coherent state transform for spaces of connections, 
{\it J. Funct. Anal.} (to be published)

\item{[4]} R. Loll: Non-perturbative solutions for lattice quantum
gravity, {\it Nucl. Phys.} B444 (1995) 619-39

\item{[5]} J.F. Barbero G.: Real Ashtekar variables for Lorentzian
signature space-times, {\it Phys. Rev.} D51 (1995) 5507-10

\item{[6]} S. Holst: Barbero's Hamiltonian derived from a
generalized Hilbert-Palatini action, Stockholm U. {\it preprint}
USITP-95-10

\item{[7]} K. Ezawa, Multi-plaquette solutions for discretized 
Ashtekar gravity, {\it Mod. Phys. Lett.} A (to be published)

\item{[8]} C. Rovelli and L. Smolin: Discreteness of area and 
volume in quantum gravity, {\it Nucl. Phys.} B442 (1995)
593-622, Err. {\it ibid.} B456 (1995) 753-4

\item{[9]} R. Loll: The volume operator in discretized quantum
gravity, {\it Phys. Rev. Lett.} 75 (1995) 3048-51

\item{[10]} R. Capovilla, T. Jacobson and J. Dell: General
Relativity without the metric, {\it Phys. Rev. Lett.} 63 (1989)
2325-8

\item{[11]} H. Kodama: Holomorphic wave function of the universe,
{\it Phys. Rev.} D42 (1990) 2548-65

\item{[12]} J.F. Barbero G.: Solving the constraints of general
relativity, {\it Class. Quant. Grav.} 12 (1995) L5-10

\item{[13]} B. Br\"ugmann, R. Gambini and J. Pullin: Jones
polynomials for intersecting knots as physical states of quantum
gravity, {\it Nucl. Phys.} B385 (1992) 587-603; How the Jones
polynomial gives rise to physical states of quantum general
relativity, {\it Gen. Rel. Grav.} 25 (1993) 1-6
 
\item{[14]} R. Loll: Spectrum of the volume operator in
quantum gravity, {\it Nucl. Phys.} B (to be published)
  
\item{[15]} R. De Pietri and C. Rovelli: Geometry eigenvalues
and scalar product from recoupling theory in loop quantum gravity,
Parma U. {\it preprint} UPRF 96-444  

\item{[16]} P. Renteln and L. Smolin: A lattice approach to spinorial
quantum gravity, {\it Class. Quant. Grav.} 6 (1989) 275-94

\item{[17]} J.B. Kogut and L. Susskind: Hamiltonian formulation of
Wilson's lattice gauge theories, {\it Phys. Rev.} D11 (1975)
395-408; J.B. Kogut: The lattice gauge theory approach
to quantum chromodynamics, {\it Rev. Mod. Phys.} 55 (1983) 775-836
 
\item{[18]} J.B. Baez: Spin network states in gauge theory,
{\it Adv. Math.} (to be published); Spin networks in nonperturbative
quantum gravity, in: {\it Proceedings of the AMS Short Course on Knots
and physics} (to be published); C. Rovelli
and L. Smolin: Spin networks and quantum gravity, {\it Phys. Rev.}
D52 (1995) 5743-59

\item{[19]} R. Loll: Independent SU(2)-loop variables and the reduced
configuration space of SU(2)-lattice gauge theory, {\it Nucl. Phys.}
B368 (1992) 121-42; Yang-Mills theory without Mandelstam
constraints, {\it Nucl. Phys.} B400 (1993) 126-44

\item{[20]} A. Ashtekar and J. Lee: Weak Field limit of general
relativity in terms of new variables: a Hamiltonian framework,
{\it Int. J. Mod. Phys.} D3 (1994) 675-94

\item{[21]} V. Husain: The $G_{\rm Newton}\rightarrow\infty$ limit
of quantum gravity, {\it Class. Quant. Grav.} 5 (1988) 575-82

\item{[22]} L. Smolin: The $G_{\rm Newton}\rightarrow 0$ limit of
Euclidean quantum gravity, {\it Class. Quant. Grav.} 9 (1992) 883-94

\item{[23]} T. Thiemann: Reality conditions inducing transforms
for quantum gauge field theory and quantum gravity, Penn State U.
{\it preprint} CGPG-95-11-4; A. Ashtekar: A generalized Wick
transform for gravity, Penn State U. {\it preprint} CGPG-95-12-1

\end